\documentclass[conference]{IEEEtran}
\IEEEoverridecommandlockouts
\usepackage{cite}
\usepackage{amsmath,amssymb,amsfonts}
\usepackage{graphicx}
\usepackage{textcomp}
\usepackage{xcolor}
\usepackage{colortbl}
\usepackage{soul}
\usepackage{booktabs}
\usepackage{multirow}
\usepackage[hidelinks]{hyperref}
\usepackage{balance}

\begin{document}
\bstctlcite{IEEEexample:BSTcontrol}

\title{Label-Budgeted Attribute Inference from Protected Features in Edge–Cloud Split Multimodal LLMs}

\author{
\IEEEauthorblockN{
Junguang Yao,
Peichun Hua,
Hanxiu Zhang,
and Yue Zheng\IEEEauthorrefmark{1}
}
\IEEEauthorblockA{
The Chinese University of Hong Kong (Shenzhen), Shenzhen, Guangdong 518172, China\\
\IEEEauthorrefmark{1}Corresponding author: zhengyue@cuhk.edu.cn
}
}

\maketitle

\begin{abstract}
Edge--cloud split inference runs the early layers of a multimodal large language model (MLLM) on a device and sends intermediate features to a server to complete the task. While keeping raw images local is often perceived as a privacy benefit, an honest-but-curious server can still infer sensitive attributes, such as age, gender, and race, from the received features. Existing protections intervene at the input, transmitted representation, or model's responses, but their native evaluations rely on final-response refusals or fixed-classifier failures, which underestimate an adaptive server's capabilities.
To address this, we propose a label-budgeted attribute-inference framework that measures representation-level privacy through a recovery curve: the share of the unprotected leakage the server recovers as a function of its labeled-sample budget, from zero (a classifier trained on unprotected features) to a full labeled set.
Our evaluation across two popular MLLMs reveals two critical findings. First, in unprotected pipelines, intermediate features leak all three attributes regardless of the split depth. Second, even when protection mechanisms are deployed across different stages, sensitive information remains recoverable under limited label budgets.
For example, Safe-LLaVA removes biometric content from the model's responses, but from the features it transmits in split inference, 20 labeled samples per class recover 75–85\% of the unprotected leakage.
This work provides a label-budgeted measure of feature-level privacy for split MLLM inference and motivates defenses that act on the transmitted representation.
\end{abstract}


\section{Introduction}

Multimodal large language models (MLLMs) answer open-ended questions about images, but their computational and memory demands make full on-device deployment difficult. Edge--cloud split inference addresses this limitation by partitioning a pretrained model between a device and a server~\cite{gupta2018distributed}: a phone, camera, or wearable runs the vision encoder and the decoder up to a chosen \emph{split layer} and transmits an intermediate representation, and a server runs the remaining layers~\cite{kang2017neurosurgeon,matsubara2023split,bajic2021collaborative,duan2020video,yuan2024split}. The transmitted representation may be a token-level \emph{hidden-state sequence} from which the server continues text generation, or a compact \emph{embedding} used for retrieval, matching, or verification~\cite{jiang2024e5v,jiang2024vlm2vec}. Compact MLLMs make the device-side computation increasingly practical~\cite{chu2023mobilevlm,yao2024minicpmv}. 

While keeping raw images local is often presented as a privacy benefit, the transmitted representation computed from the entire image may still reveal sensitive information to the server. Learned representations can encode attributes unrelated to the intended task~\cite{song2020overlearning,melis2019exploiting}, and embeddings may leak sensitive attributes or support input inversion~\cite{song2020information,pan2020privacy,morris2023text,feng2026inverting}. We therefore study \emph{attribute inference}, in which an honest-but-curious server follows the inference protocol but additionally attempts to infer sensitive attributes (e.g., age, gender, and race from face images) from the representations it legitimately receives. 
% These risks motivate protecting representations shared with a server~\cite{hua2026pointing}.
% and a curious server can reconstruct the inputs behind split features~\cite{he2019model,pasquini2021unleashing}. 

Defenses against this leakage exist for CNN split inference. DeepObfuscator~\cite{li2021deepobfuscator}, TIPRDC~\cite{li2020tiprdc}, and DISCO~\cite{singh2021disco} train or prune the device-side network against an attribute adversary; Shredder~\cite{mireshghallah2020shredder} and NoPeek~\cite{vepakomma2020nopeek} add learned noise or decorrelate the features from the input; Roulette~\cite{li2024roulette} hides the class label. These methods are designed for closed-set classification tasks. In an MLLM, by contrast, the server-side decoder must generate open-ended text from the transmitted features, and, to our knowledge, no attribute defense has been designed for this setting. A deployer may instead adapt existing defenses built for other threat models. These mechanisms intervene at different points in the inference pipeline or provide protection at different interfaces (Fig.~\ref{fig:system}). At the \emph{input level}, ImageProtector~\cite{shao2026imageprotector} adds a near-imperceptible perturbation that makes the model refuse to analyze the image. At the \emph{representation level}, PASS~\cite{dhar2021pass} transforms a face embedding so that a retrained attribute classifier fails while face verification still works. Safe-LLaVA~\cite{kim2025safellava} is a model-level intervention with an \emph{output-level} protection objective: it retrains LLaVA on data with biometric information removed, so that the model refuses biometric queries and reduces ``implicit biometric leakage in general responses''. Split inference, however, exposes intermediate features to the server before any response is generated. Protection therefore has to be measured on those features. The native evaluations of the above methods examine different outcomes, such as model refusal or the accuracy of a particular attribute classifier. We argue that these only cover the two ends of a curve: the server's accuracy as a function of how many labeled protected samples it holds.
% Existing evaluations test one classifier, trained either on unprotected features or on a full labeled set of protected ones, and report one accuracy. We argue that these only cover the two ends of a curve: the server's accuracy as a function of how many labeled protected samples it holds.

\textbf{We therefore ask: how much data from the protected domain does a server need to recover a sensitive attribute from protected features?} We answer with an attribute-inference framework whose single knob is the attacker's budget $b$ of labeled protected features per class. At $b=0$ the attacker is a classifier trained on unprotected features (the \emph{transfer} attacker); with $b\le20$ it fine-tunes that classifier using a few labeled protected features (\emph{few-label}); with a full protected auxiliary set it retrains the classifier (\emph{adaptive}). 

Using face images in generation and face-recognition settings, we study the recovery of age, gender, and race. Our framework characterizes leakage through a \emph{recovery curve} that quantifies the share of unprotected leakage recovered under each budget of labeled protected features, complemented by a classifier-free dependence test (Sections~\ref{sec:model}--\ref{sec:audit}). Our measurements on a device--server GPU pair find that all three attributes remain recoverable at every tested split layer of both models, while deeper splits raise device cost without meaningfully reducing measured attribute leakage (Section~\ref{sec:results}). Our evaluation further shows that the native metrics of protection mechanisms with different intervention points and security objectives do not reliably indicate privacy at the split boundary. Safe-LLaVA's model-level protection keeps the attributes in its features, where 20 labels per class recover 75--85\% of the unprotected leakage; the input-level ImageProtector protects only the model it was optimized for; and the representation-level PASS is more resistant to the evaluated attacks but retains detectable statistical dependence (Sections~\ref{sec:transfer}--\ref{sec:fewlabel}).

% Our contributions are:
% \begin{itemize}
% \item \textit{Capability-aware framework:} an attribute-inference framework for MLLM split inference whose \emph{recovery curve} quantifies the share of unprotected leakage recovered under each budget of labeled protected representations, complemented by a classifier-free dependence test (Sections~\ref{sec:model}--\ref{sec:audit}).
% \item \textit{Split-layer analysis:} measurements on a client--server GPU pair showing that all three attributes remain recoverable at every tested split layer of both models, while deeper splits raise device cost without meaningfully reducing measured attribute leakage (Section~\ref{sec:depth}).
% \item \textit{Cross-mechanism privacy analysis:} a unified evaluation showing that the native metrics of protection mechanisms with different intervention points and security objectives do not reliably indicate privacy at the split boundary. Safe-LLaVA's output-level protection keeps the attributes in its features, where 20 labels per class recover 75--85\% of the unprotected leakage; the input-level ImageProtector protects only the model it was optimized for; and the representation-level PASS is more resistant to the evaluated attacks but retains detectable statistical dependence (Sections~\ref{sec:native}--\ref{sec:fewlabel}).
% \end{itemize}

\section{System and Threat Model}\label{sec:model}

\begin{figure}[t]
\centering
\includegraphics[width=\columnwidth]{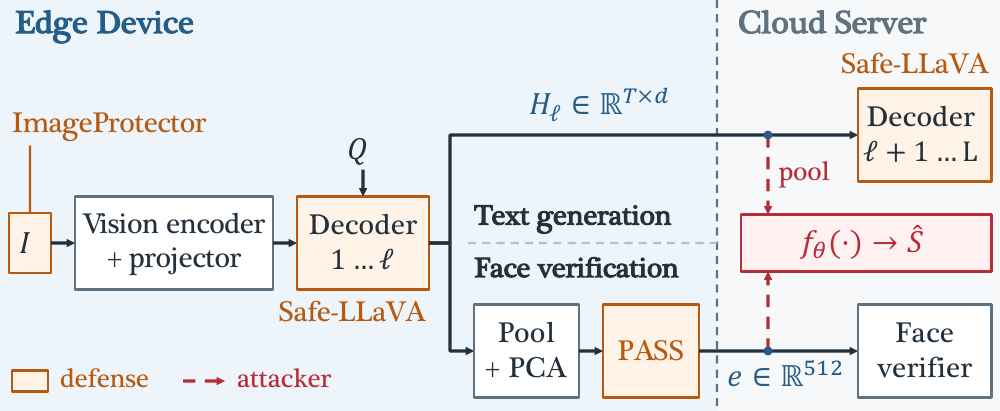}
\caption{Split inference for text generation and face verification. Orange marks what each defense changes: the image (ImageProtector), the decoder on both sides of the split (Safe-LLaVA), and the embedding (PASS).}
\label{fig:system}
\end{figure}

\emph{System and Parties.} We consider a provider-owned split-inference service (Fig.~\ref{fig:system}). The service provider owns the complete MLLM and deploys its device-side partition to the user's trusted device. Given an image $I$ and prompt $Q$, the device runs the vision encoder, the projector, and decoder layers $1..\ell$ of $L$, including any prescribed protection mechanism. The provider's server receives the resulting features and completes the task: it runs layers $\ell{+}1..L$ to generate a response, or uses a compact embedding for face verification. 

\emph{Transmitted Features.} For generation, the device sends
\begin{equation}
H_\ell(I,Q) = [h_{\ell,1},\dots,h_{\ell,T}] \in \mathbb{R}^{T\times d},
\end{equation}
where the token count $T$ and width $d$ do not depend on $\ell$. For recognition, the device pools $H_\ell$ over tokens and projects it to an embedding $e$, which the server uses for face verification. We call features produced without and with a protection mechanism \emph{unprotected} and \emph{protected features}, respectively; a \emph{protected sample} is a protected feature with its attribute label.
% 2.48\,MB per image in FP16 for LLaVA-1.5-7B ($T=303$) and 0.36\,MB for Qwen2.5-VL-3B ($T=89$ on FairFace)
% the device pools $H_l$ over tokens and projects it to a 512-dimensional embedding $e$ (1\,KB in FP16)

\emph{Adversary.} The server is honest-but-curious: it faithfully completes the task but tries to infer an unrequested sensitive attribute $S\in\{\text{age},\text{gender},\text{race}\}$ from the received features. As the service provider, it knows the architecture, the split layer, and the protection mechanism, and holds a public auxiliary face set $\mathcal{D}_{\mathrm{aux}}$ whose sensitive-attribute labels are available from dataset annotations or other metadata. It can locally execute the device pipeline on $\mathcal{D}_{\mathrm{aux}}$ to generate matching unprotected or protected auxiliary features. 
% Because the provider owns and deploys the device-side partition, it can also run the same client pipeline locally on auxiliary images. Passing $\mathcal{D}_{\mathrm{aux}}$ through the unprotected or protected device-side pipeline provides the corresponding feature representations; the attribute labels are obtained independently of the model.
The attacker trains a probe $f_\theta$, a classifier or regressor, to predict $S$~\cite{alain2017understanding,belinkov2022probing}. Its budget $b$ counts the protected samples it holds, per class for categorical attributes: at $b=0$, it transfers a probe trained on unprotected features; for $b\leq 20$, it adapts the probe on a few labeled samples from users whose attributes it already knows, e.g., from account profiles; at full budget, it uses the complete protected auxiliary set. From a sequence, the attacker mean-pools the features, $z_\ell=\frac{1}{T}\sum_{t} h_{\ell,t}$; from an embedding, it uses $e$ directly.

\section{Attribute-Inference Framework}\label{sec:audit}
We propose a label-budgeted attribute-inference framework that formulates the attacker's strength as a budget $b$ of protected samples. Auxiliary images are disjoint from the target images, those of the users under attack, and no attacker alters the device pipeline or sees target labels. The \emph{transfer} attacker ($b=0$) trains $f_\theta$ on unprotected auxiliary features and applies it directly to protected target features; the \emph{few-label} attacker fine-tunes it on $b\in\{1,2,5,10,20\}$ protected samples per class (FT); a transductive variant additionally self-trains on unlabeled protected target features (ST). Finally, the \emph{adaptive} attacker trains $f_\theta$ on the full protected auxiliary set to establish an empirical upper bound on leakage.
% it is what a designer obtains by reusing an existing attribute classifier.

Evaluating privacy leakage across different attributes poses a challenge due to varying metrics and baselines. Categorical attributes (gender, race, and age group) are evaluated using balanced accuracy (BA)~\cite{brodersen2010balanced}, where chance is $1/K$ for $K$ classes, while the continuous regression task (age in years) is evaluated by mean absolute error (MAE). Because BA and MAE are not comparable, we propose a \emph{recovery curve} $R(b)$ to normalize these diverse tasks relative to the task baseline and unprotected leakage. The \emph{recovery} at budget $b$ is
\begin{equation}
R(b) = \frac{m(b)-m_0}{m_{\mathrm{unp}}-m_0},
\end{equation}
where $m(b)$ is the attacker's BA or negative MAE at budget $b$, $m_0$ is the task-specific baseline (chance accuracy for classification or mean-predictor baseline for regression), and $m_{\mathrm{unp}}$ is the score of a probe trained and tested on unprotected features. A mechanism that merely shifts the features~\cite{elazar2018adversarial} yields a low $R(0)$ and a curve that rises quickly with $b$; attribute removal keeps $R$ near zero even at the full budget. Values above 100\% exceed the unprotected leakage. We also report the \emph{gap closed}, $(R(b)-R(0))/(R(\mathrm{full})-R(0))$: 0\% when $b$ labels do no better than the transfer attacker, and 100\% when they match the adaptive one. 
% the share of what full retraining adds that $b$ labels already buy.

\emph{Dependence Test.} A failed probe does not guarantee privacy. We therefore also test, without training any classifier, whether the protected features still depend statistically on the attribute. The test statistic is the normalized Hilbert--Schmidt independence criterion (nHSIC)~\cite{gretton2005measuring} between protected target features and the attribute. We use a Gaussian kernel on the features (median-heuristic bandwidth) and a delta or Gaussian kernel for categorical or continuous labels, and estimate the $p$-value from 1{,}000 label permutations.

\section{Experimental Setup}\label{sec:setup}
% \emph{Utility.} Each defense keeps the utility metric of its setting (Table~\ref{tab:native}): for generation, the CLIP image--text similarity between the unperturbed image and the response to a benign prompt, relative to the model without the defense; for recognition, face verification accuracy on the 6{,}000-pair LFW protocol~\cite{huang2007labeled}.

We evaluate LLaVA-1.5-7B~\cite{liu2024improved} (LLaVA; 32 decoder layers, $d=4096$) and Qwen2.5-VL-3B~\cite{bai2025qwen25vl} (Qwen; 36 layers, $d=2048$) using the generation prompt ``Describe this image''. We use FairFace (FF)~\cite{karkkainen2021fairface} to evaluate categorical age (9 groups), gender (2), and race (7), yielding a chance BA of 0.11, 0.50, and 0.14, respectively. The aligned and cropped UTKFace (UTK)~\cite{zhang2017age} (23{,}705 images) is used to evaluate gender, race (5 categories, chance 0.20), and age in years. For UTK age, the mean-predictor baseline MAE is 15.8 years.

% after layer $l\in\{5,15,25,L\}$. FairFace (FF)~\cite{karkkainen2021fairface} labels age (9 groups), gender (2), and race (7); the aligned and cropped UTKFace (UTK)~\cite{zhang2017age} (23{,}705 images) labels gender (2), race (5), and age in years. Chance BA is thus 0.11, 0.50, and 0.14 for FF age, gender, and race, and 0.20 for UTK race. The prompt is ``Describe this image.''
To evaluate the privacy--cost tradeoff, models are split after layer $\ell\in\{5,15,25,L\}$, with an RTX 2080 Ti (11\,GiB) as the device and a V100 (32\,GiB) as the server. The transmitted sequence is 2.48\,MB per image in FP16 for LLaVA ($T=303$) and 0.36\,MB for Qwen ($T=89$ on FF); the embedding $e$ has 512 dimensions (1\,KB). We report device (Dev.) latency, peak GPU memory, and energy above idle power, and server (Srv.) latency; latencies are medians in ms of 100 FP16 runs after a 10-run warmup. 

We evaluate ImageProtector by attribute-query refusal on FF/UTK, Safe-LLaVA by soft/hard refusal on the PRISM benchmark~\cite{kim2025safellava}, and PASS by attribute BA on FF. Safe-LLaVA uses the released 7B checkpoint on both sides of the split. ImageProtector~\cite{shao2026imageprotector} optimizes the perturbation through LLaVA with 1{,}000 steps of projected iterative FGSM~\cite{goodfellow2015explaining} ($\ell_\infty$ budget $8/255$, step 0.007); on Qwen we feed the same perturbed images, which tests cross-model transfer. PASS~\cite{dhar2021pass} is trained on the 512-dimensional layer-15 embedding of each model, one per attribute. Utility for Safe-LLaVA and ImageProtector is the CLIP~\cite{radford2021learning} image--text similarity between the unperturbed image and the response to a benign prompt, relative to the unprotected model; for PASS it is face-verification accuracy on the 6{,}000-pair Labeled Faces in the Wild (LFW) protocol~\cite{huang2007labeled}.

\emph{Probes.} Final-layer probes use an 80/20 stratified split of each dataset; all other evaluations use 1{,}050 auxiliary ($\mathcal{D}_{\mathrm{aux}}$) and 2{,}100 target images and report the mean and standard deviation over five seeds. Few-label attackers self-train on one half of the target set and are tested, with the transfer and adaptive attackers, on the other. The split-depth study uses a linear probe. The protection evaluation uses layer-15 features, standardized, projected to 512 dimensions by PCA, and standardized again, with all three steps fitted on unprotected auxiliary features and then fixed; PASS uses its own embedding. A fixed MLP (256 hidden units, dropout 0.1; AdamW, learning rate $10^{-3}$, batch size 128) predicts categorical attributes, and ridge regression ($\alpha=1$) predicts age.

% \emph{Deployment.} The client is an RTX 2080 Ti (11\,GB) that holds only the vision encoder and decoder layers $1..l$; the server is a V100 (32\,GB). On FF inputs we measure the client (Dev.) latency, peak GPU memory, and energy per image, and the server (Srv.) latency from the split to the last layer. Latencies are medians of 100 FP16 runs after 10 warm-up runs, and energy integrates sampled GPU power above idle.

\section{Results}\label{sec:results}

% \subsection{Unprotected Features Leak at Every Split Layer}\label{sec:depth}
\begin{table}[t]
\centering
\caption{Unprotected BA (MAE for UTK age) at each split layer.}
\label{tab:depth}
\setlength{\tabcolsep}{3.2pt}
\begin{tabular}{llccccc}
\toprule
Model & Attribute & Baseline & $\ell=5$ & $\ell=15$ & $\ell=25$ & Last \\
\midrule
\multirow{6}{*}{LLaVA}
& FF age & 0.111 & 0.554 & 0.556 & 0.558 & 0.558 \\
& FF gender & 0.500 & 0.955 & 0.956 & 0.954 & 0.953 \\
& FF race & 0.143 & 0.691 & 0.697 & 0.683 & 0.677 \\
& UTK gender & 0.500 & 0.954 & 0.955 & 0.953 & 0.953 \\
& UTK race & 0.200 & 0.791 & 0.796 & 0.784 & 0.782 \\
& UTK age (MAE) & 15.8 & 6.75 & 7.23 & 6.77 & 7.23 \\
\midrule
\multirow{6}{*}{Qwen}
& FF age & 0.111 & 0.539 & 0.531 & 0.542 & 0.515 \\
& FF gender & 0.500 & 0.915 & 0.913 & 0.909 & 0.903 \\
& FF race & 0.143 & 0.634 & 0.634 & 0.629 & 0.622 \\
& UTK gender & 0.500 & 0.941 & 0.937 & 0.935 & 0.937 \\
& UTK race & 0.200 & 0.754 & 0.748 & 0.750 & 0.749 \\
& UTK age (MAE) & 15.8 & 6.75 & 6.76 & 6.90 & 6.82 \\
\bottomrule
\end{tabular}
\end{table}

\begin{table}[t]
\centering
\caption{Cost of each split layer (latency in ms). OOM: out of memory.}
\label{tab:cost}
\setlength{\tabcolsep}{3pt}
\begin{tabular}{lcccccccc}
\toprule
 & \multicolumn{4}{c}{LLaVA-1.5-7B} & \multicolumn{4}{c}{Qwen2.5-VL-3B} \\
\cmidrule(lr){2-5}\cmidrule(lr){6-9}
$\ell$ & Dev. & GiB & J & Srv. & Dev. & GiB & J & Srv. \\
\midrule
5 & 35.6 & 2.78 & 4.6 & 73.4 & 47.0 & 2.58 & 3.2 & 53.5 \\
15 & 71.8 & 6.57 & 10.1 & 46.7 & 61.5 & 4.05 & 4.4 & 30.4 \\
25 & \multicolumn{3}{c}{OOM} & 20.1 & 73.7 & 5.51 & 5.5 & 16.8 \\
Last & \multicolumn{3}{c}{OOM} & 1.4 & 89.7 & 7.13 & 6.8 & 2.3 \\
\bottomrule
\end{tabular}
\end{table}

On the full 80/20 dataset splits, final-layer probes recover all evaluated attributes significantly better than baseline predictors. Table~\ref{tab:depth} shows that this leakage does not fade with depth: from layer 5 to the last layer, no BA moves by more than 0.03, and age MAE by at most 0.5 years. The cost does change (Table~\ref{tab:cost}): moving the LLaVA split from layer 5 to layer 15 doubles device latency and energy, and layer 25 no longer fits the 11-GiB device. Sending the LLaVA sequence takes 199\,ms at 100\,Mbit/s, longer than the device compute at any split that fits. \emph{Moving the split deeper raises device cost without meaningfully reducing measured attribute leakage.}

\begin{figure*}[t]
\centering
\includegraphics[width=\textwidth]{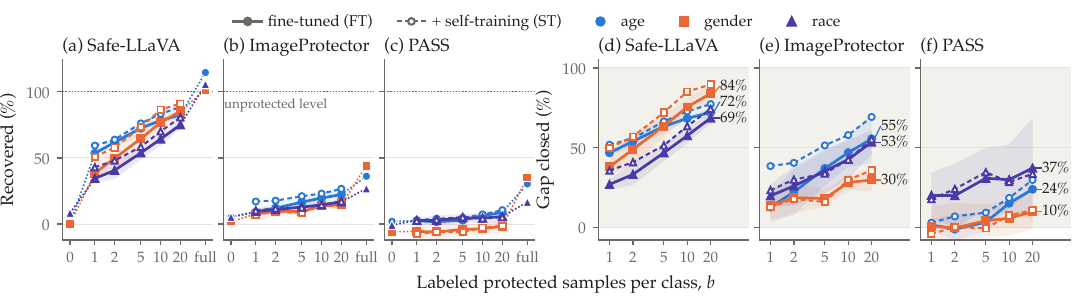}
\caption{Recovery curves on FF with LLaVA: recovery $R(b)$ in (a)--(c) and gap closed in (d)--(f) against the label budget.}
\label{fig:fewlabel}
\end{figure*}

% \subsection{Each Defense Passes Its Own Test}\label{sec:native}

\begin{table}[t]
\centering
\caption{Each defense's own protection and utility results, by level.}
\label{tab:native}
\setlength{\tabcolsep}{1.6pt}
\begin{tabular}{@{}lll@{}}
\toprule
Defense & Protection & Utility \\
\midrule
\multicolumn{3}{l}{\textbf{\emph{Input level}}} \\
ImageProt. & Refusal 97.6--99.6\% (FF/UTK) & CLIP 98.5\% FF, 100.4\% UTK \\
 & & PSNR 34.0\,dB, SSIM 0.83 \\
\multicolumn{3}{l}{\textbf{\emph{Model level}}} \\
Safe-LLaVA & PRISM refusal 98.5\% / 96.3\% & CLIP 94.0\% FF, 91.4\% UTK \\
\multicolumn{3}{l}{\textbf{\emph{Representation level} (FF; LLaVA / Qwen)}} \\
PASS-age & BA .221 (.484) / .177 (.457) & LFW .876 / .827 \\
PASS-gender & BA .608 (.805) / .539 (.783) & LFW .839 / .870 \\
PASS-race & BA .200 (.555) / .207 (.499) & LFW .803 / .810 \\
\bottomrule
\end{tabular}
\end{table}

We verify that each protection mechanism fulfills its native security objective (Table~\ref{tab:native}). Safe-LLaVA refuses biometric queries on PRISM and keeps over 90\% of the unprotected CLIP score. ImageProtector makes LLaVA refuse the attribute queries it was optimized against and still answer the benign prompt, but protecting one image takes 295\,s and 17.6\,GiB on the V100, more than the 11-GiB device holds. PASS lowers its retrained attribute classifier's accuracy toward chance (unprotected BA in parentheses) while keeping 88--99\% of unprotected face-verification accuracy.

\subsection{Recovery at Zero and Full Budget}\label{sec:transfer}

\begin{table}[t]
\centering
\caption{Leakage (BA; MAE in years for UTK age) at zero ($b{=}0$) and full budget, shaded by recovery $R$ (full red: 100\%). Last column: protected over unprotected nHSIC.}
\label{tab:transfer}
\setlength{\tabcolsep}{2.6pt}
\begin{tabular}{llccccccc}
\toprule
 & & \multicolumn{3}{c}{FF} & \multicolumn{2}{c}{UTK} & UTK & nHSIC \\
\cmidrule(lr){3-5}\cmidrule(lr){6-7}
Defense & & Age & Gen. & Race & Gen. & Race & Age & ratio \\
\midrule
\multicolumn{9}{l}{\emph{Generation: transmitted sequence}} \\
Safe-LLaVA & Unprot. & .496 & .940 & .637 & .943 & .778 & 5.79 & \multirow{3}{*}{2.2--7.3} \\
(LLaVA) & $b{=}0$ & \cellcolor{red!9}.181 & \cellcolor{red!7}.563 & \cellcolor{red!7}.208 & \cellcolor{red!25}.722 & \cellcolor{red!10}.315 & 35.58 & \\
 & Full & \cellcolor{red!50}.552 & \cellcolor{red!50}.954 & \cellcolor{red!50}.690 & \cellcolor{red!50}.957 & \cellcolor{red!50}.801 & \cellcolor{red!45}6.73 & \\
\cmidrule(l){2-9}
ImageProt. & Unprot. & .455 & .849 & .562 & .915 & .718 & 6.25 & \multirow{3}{*}{0.78--2.3} \\
(Qwen) & $b{=}0$ & \cellcolor{red!33}.339 & \cellcolor{red!41}.789 & \cellcolor{red!28}.378 & \cellcolor{red!36}.802 & \cellcolor{red!31}.517 & \cellcolor{red!32}9.61 & \\
 & Full & \cellcolor{red!50}.456 & \cellcolor{red!50}.846 & \cellcolor{red!44}.515 & \cellcolor{red!50}.916 & \cellcolor{red!49}.710 & \cellcolor{red!41}8.00 & \\
\cmidrule(l){2-9}
ImageProt. & Unprot. & .496 & .940 & .637 & .943 & .778 & 5.79 & \multirow{3}{*}{0.29--0.94} \\
(LLaVA) & $b{=}0$ & \cellcolor{red!5}.151 & \cellcolor{red!16}.637 & \cellcolor{red!3}.170 & \cellcolor{red!24}.713 & \cellcolor{red!8}.293 & \cellcolor{red!7}14.45 & \\
 & Full & \cellcolor{red!22}.279 & \cellcolor{red!23}.703 & \cellcolor{red!15}.291 & \cellcolor{red!34}.805 & \cellcolor{red!22}.449 & \cellcolor{red!13}13.28 & \\
\midrule
\multicolumn{9}{l}{\emph{Recognition: transmitted embedding}} \\
PASS & Unprot. & .484 & .805 & .555 & .846 & .672 & 6.73 & \multirow{3}{*}{0.17--0.43} \\
(LLaVA) & $b{=}0$ & .116 & \cellcolor{red!5}.530 & .146 & .510 & .210 & 15.71 & \\
 & Full & \cellcolor{red!15}.221 & \cellcolor{red!18}.608 & \cellcolor{red!7}.200 & \cellcolor{red!19}.634 & \cellcolor{red!16}.354 & 15.52 & \\
\cmidrule(l){2-9}
PASS & Unprot. & .457 & .783 & .499 & .823 & .641 & 7.29 & \multirow{3}{*}{0.14--0.30} \\
(Qwen) & $b{=}0$ & .117 & .514 & .154 & \cellcolor{red!4}.529 & .196 & 15.65 & \\
 & Full & \cellcolor{red!10}.177 & \cellcolor{red!7}.539 & \cellcolor{red!9}.207 & \cellcolor{red!12}.578 & \cellcolor{red!11}.294 & 15.53 & \\
\bottomrule
\end{tabular}
\end{table}

Under Safe-LLaVA, ImageProtector on LLaVA, and PASS, the transfer attacker recovers little of the unprotected leakage (Table~\ref{tab:transfer}), so a check with an existing classifier would suggest that all three are protected. The full budget separates them by level. Safe-LLaVA's model-level protection retains every attribute in its features: adaptive recovery $R$ reaches 91--115\% of the unprotected leakage, ranging from UTK age (MAE 6.73 vs. 5.79) to FF age (BA 0.552 vs. 0.496), and nHSIC is \mbox{2--7} times the unprotected baseline ($p<0.001$). \emph{The transfer attacker's failure is the failure of an old classifier, not the removal of the attribute.} The input-level ImageProtector limits adaptive recovery $R$ to 25--69\% on LLaVA, ranging from UTK age (MAE 13.28 vs. 5.79) to UTK gender (BA 0.805 vs. 0.943); however, it provides little cross-model protection: on Qwen, even the transfer attacker recovers most of the leakage, and the adaptive attacker recovers 82--100\%, ranging from UTK age (MAE 8.00 vs. 6.25) to approximately full recovery on FF age and UTK gender. The representation-level PASS caps the adaptive attacker at 39\%, with the maximum occurring for UTK gender (BA 0.634 vs. 0.846); UTK age remains near the mean-predictor baseline (15.52 vs. 15.8 years), consistent with its native evaluation. Its nHSIC, however, remains significant in every condition ($p\le0.033$), so a dependence the probes do not exploit survives in the embedding.

\subsection{Recovery Curves}\label{sec:fewlabel}

Fig.~\ref{fig:fewlabel} plots $R(b)$ in (a)--(c) and the gap closed in (d)--(f) on FF with LLaVA. Dashed curves add self-training (ST) to fine-tuning (FT), and shaded bands show one standard deviation. Under Safe-LLaVA, one protected sample per class already recovers 35--53\% of the unprotected leakage. Twenty samples per class, that is, 40 labeled images for gender, 140 for race, and 180 for age, recover 75--85\% and close 69--84\% of the gap. \emph{Recovering half of the unprotected leakage costs the server 1 to 5 labels per class under Safe-LLaVA; on FF, it remains out of reach under ImageProtector and PASS even with the full set.} Twenty labels per class recover at most about 27\% under ImageProtector and 10\% under PASS. For UTK age under Safe-LLaVA, a separate residual-ridge calibration lowers the transfer attacker's MAE from 35.3 years to 13.3 with 10 protected samples and to 9.2 with 200, approaching 6.8 at the full budget.
%Fig.~\ref{fig:fewlabel}(d) scores stronger probes against their default: replacing all-token mean-pooling with non-visual token positions or learned token fusion recovers up to 56\% more leakage on LLaVA and about 5\% more under Safe-LLaVA, and adding layers 15 and 25 to a last-layer probe recovers 35\% more UTK age leakage on Qwen than the last layer alone.

% \subsection{Robustness of the Leakage}\label{sec:robust}

% Our one-layer, mean-pooled probe may understate what the server can learn. Fig.~\ref{fig:fewlabel}(d) scores stronger probes against it on the same features. Using non-visual token positions or learned token fusion recovers up to 56\% more leakage on LLaVA and 8\% more under Safe-LLaVA. Combining layer $l=15$ with layer 25 and the last recovers 7\% more UTK age leakage on LLaVA and 35\% more on Qwen. Three equivalent benign prompts change BA by less than 0.01.

\section{Conclusion}\label{sec:conclusion}
The attribute-level privacy of split MLLM inference depends on where the protection mechanism intervenes. Safe-LLaVA's model-level protection leaves sensitive attributes recoverable from intermediate features, requiring only 1 to 5 labels per class to recover half of the unprotected leakage. The input-level ImageProtector provides limited cross-model protection, while the representation-level PASS suppresses inference attacks but retains measurable dependence. Moving the split layer deeper increases device overhead without mitigating leakage. Privacy mechanisms for split inference should be evaluated using label budgets and classifier-free dependence tests. Extending the framework beyond faces and to compressed features remains future work.

\clearpage
\balance
\bibliographystyle{IEEEtran}
\bibliography{refs}

\end{document}